\documentclass[12pt,a4paper]{article}
\usepackage{latexsym}
\usepackage[dvips]{graphics}
\usepackage{epsfig}
\usepackage{color}
\usepackage{amsmath,amssymb}
\usepackage{cite,mcite}
\usepackage{xspace}
\usepackage{ifthen}
\usepackage{hyperref}
\usepackage{slashed}
\usepackage{epstopdf}
\usepackage[rightcaption]{sidecap}
\sidecaptionvpos{figure}{c}
\usepackage[hang,small]{caption}
\usepackage{here}
\usepackage{placeins}

\usepackage{tabularx}

\newcommand{\beq}{\begin{equation}}
\newcommand{\eeq}{\end{equation}}
\newcommand{\bea}{\begin{eqnarray}}
\newcommand{\eea}{\end{eqnarray}}

\newcommand{\be}{\begin{eqnarray*}}
\newcommand{\ee}{\end{eqnarray*}}

\newcommand{\spin}[1]{\mbox{spin-#1}}
\newcommand{\ff}{\ensuremath\mathit{ff}}

\newcolumntype{C}[1]{>{\centering\arraybackslash}p{#1}}

\numberwithin{equation}{section}
\begin{document}

\thispagestyle{empty}

\begin{center}
\hfill KA-TP-10-2013
\end{center}

\begin{center}

\vspace{1.7cm}

{\LARGE\bf 
Higgs Spin Determination in the WW channel and beyond} 

\vspace{1.4cm}

{\bf Jessica Frank, 
\bf Michael Rauch, 
\bf Dieter Zeppenfeld}\\ 

\vspace{1.2cm}

{\em {Institute for Theoretical Physics, Karlsruhe 
Institute of Technology, 76128 Karlsruhe, Germany}
}\\

\end{center}

\vfill

\centerline{\bf Abstract}
\vspace{2 mm}
\begin{quote}
\small
After the discovery of the 126 GeV resonance at the LHC, the determination of 
its features, including 
its spin, is a very important ongoing task. In order to distinguish the two most 
likely spin hypotheses, \spin0 or \spin2, 
 we study the phenomenology of a
light Higgs-like \spin2 resonance produced in different gluon-fusion and 
vector-boson-fusion processes at the LHC. 
Starting from an effective model for the 
interaction of a \spin2 particle with the SM gauge bosons, we calculate cross 
sections and differential distributions within the Monte
Carlo program \textsc{Vbfnlo}. We find that with 
specific model parameters such a \spin2 resonance can mimic 
SM Higgs rates and transverse-momentum distributions 
in $\gamma \gamma$, $WW$ and $ZZ$ decays, whereas several distributions 
allow to separate \spin2 from \spin0, independently of the \spin2 model parameters.
\end{quote}

\vfill

\newpage
%
%%%%%%%%%%%%%%%%%%%%%%%%%%%%%%%%%%%%%%%%%%%%%%%%%%%%%%%
\section{Introduction \label{sec:Intro}}
%%%%%%%%%%%%%%%%%%%%%%%%%%%%%%%%%%%%%%%%%%%%%%%%%%%%%%%

A new, Higgs-like resonance with a mass of about 126 GeV has been 
discovered by the LHC experiments ATLAS~\cite{ATL-CONF-comb} and CMS~\cite{CMS-CONF-comb}. 
A very important ongoing task is now to probe whether this resonance is the Standard Model (SM) 
Higgs boson~\cite{Higgs} responsible for electroweak symmetry breaking. 
Therefore, the determination of its features like its
couplings (including its
self-couplings), CP quantum number and 
spin, is currently a very active field of research.

So far, all experimental results suggest that it is indeed the SM Higgs 
that has been found: 
The coupling strengths, measured from the rates in different decay channels at ATLAS and CMS, 
are in good agreement with the SM expectations~\cite{couplATLAS, couplCMS}.  
A pure CP odd state is already strongly disfavored~\cite{CMS_CP, LHC_spinCP_ZZ}, and also recent approaches to determine 
the spin favor \spin0 over specific \spin2 scenarios~\cite{couplCMS, LHC_spinCP_ZZ, LHC_spin_WW, ATLAS_spin_gammagamma}. 
The observation of the resonance in the di-photon decay mode immediately
excludes a \spin1 particle due to the Landau-Yang
theorem~\cite{Landau-Yang}, leaving \spin2 as an alternative hypothesis 
to the \spin0 of the Higgs boson. Due to the variety of possible spin-2 models and 
discriminating variables, there are 
many phenomenological attempts to distinguish between \spin0 and
\spin2~\cite{Frank:2012wh, spin2pheno, EllisGraviton, invdilepmass}. 

Neglecting P-wave quark-antiquark collisions, a \spin2 resonance can be produced mainly 
in gluon fusion or vector-boson fusion (VBF), which are also the most important Higgs 
production channels. The main decay modes for the observation and analysis 
of the new resonance include $\gamma \gamma \,$, $W^+ W^- \rightarrow 2 l 2 \nu$ 
and $Z Z \rightarrow 4 l$. 
Whereas in a previous analysis~\cite{Frank:2012wh} we have studied the phenomenology of 
\spin2 resonances in the VBF photon pair-production mode, we now present
 cross sections for \spin2 resonances in all these processes and show that 
with a suitable choice 
of model parameters, spin-2 resonances can approximately reproduce  
SM Higgs cross sections in all the different channels. 
We then focus on the phenomenology of \spin2 
resonances in the $WW$ channel produced in gluon fusion or VBF and 
discuss differential distributions, which can be useful to 
distinguish between a SM Higgs and a \spin2 resonance. 
Furthermore, alternative \spin0 scenarios 
are also considered. 
Our calculations are performed with the Monte Carlo program 
\textsc{Vbfnlo}~\cite{Arnold:2011wj} by using an 
effective Lagrangian approach for a \spin2 (or \spin0) field interacting with the SM  
gauge bosons. These parametrizations are presented in Section~\ref{sec:spin2model}. 
After sketching the relevant aspects of our calculation in
Section~\ref{sec:Calculation}, we present the results of our analysis in Section~\ref{sec:Results}.

%%%%%%%%%%%%%%%%%%%%%%%%%%%%%%%%%%%%%%%%%%%%%%%%%%%%%%%
 \section{\label{sec:spin2model} Spin-2 and Spin-0 Parametrization}
%%%%%%%%%%%%%%%%%%%%%%%%%%%%%%%%%%%%%%%%%%%%%%%%%%%%%%%

For the analysis of \spin2 resonances in gluon-fusion and 
vector-boson-fusion
processes, we start from an effective Lagrangian ansatz for a \spin2 singlet state 
which we have already introduced in Ref.~\cite{Frank:2012wh}. There, we have restricted ourselves to 
a model for the interaction of a \spin2 particle with electroweak
bosons, since only electroweak-boson fusion was studied. 
In order to consider also gluon fusion, we enlarge this model by a new term describing 
the gluonic interaction, 
$\frac{f_9}{\Lambda} T_{\mu \nu} G^{\alpha \nu}_{a} G^{a, \mu}_{\hspace{0.3 cm} \alpha}$, 
and end up with the effective Lagrangian
\begin{align}
 \mathcal{L}_{\text{Spin-2}}&=\frac{1}{\Lambda} T_{\mu \nu} \left(f_1 B^{\alpha \nu} B^\mu_{\hspace{0.15 cm} \alpha} 
+f_2 W_i^{\alpha \nu} W^{i, \mu}_{\hspace{0.3 cm} \alpha}+ 2f_5 (D^\mu\Phi)^\dagger(D^\nu\Phi) 
+ f_9 \, G^{\alpha \nu}_{a} G^{a, \mu}_{\hspace{0.3 cm} \alpha} \right) \label{spin2lagrangian}.
\end{align}
${\Lambda}$ is the characteristic energy scale of the underlying new
physics, $f_i$ are variable coupling parameters, $B^{\alpha \nu}$, 
$W_i^{\alpha \nu}$ and $G^{\alpha \nu}_{a}$ are the field strength tensors of the 
SM gauge bosons and
$D^\mu$ is the covariant derivative

\begin{equation}
D^\mu=\partial^\mu-igW_i^\mu \frac{\sigma^i}{2} -ig'Y B^\mu.
\end{equation}
$\Phi$ is a scalar doublet field with vacuum expectation value $v/\sqrt{2} = 174$ GeV. 
The mass of the \spin2 particle is taken as a free parameter.

The Lagrangian
(\ref{spin2lagrangian}) yields five relevant vertices, which involve two gauge bosons and the
\spin2 particle $T$, namely $TW^+W^-$, $TZZ$, $T\gamma \gamma$, $T\gamma Z$ and $Tgg$. 
The corresponding Feynman rules are:

\begin{align}
TW^+W^- &:\,\, \frac{2i f_2}{\Lambda} K_1^{\alpha \beta \mu \nu} +\frac{if_5 g^2 v^2}{2 \Lambda} K_2^{\alpha \beta \mu \nu}, \nonumber\\
TZZ &:\,\, \frac{2i}{\Lambda} (f_2 c_w^2+f_1 s_w^2) K_1^{\alpha \beta \mu \nu} + \frac{if_5 v^2}{2 \Lambda} (g^2+g'^2) K_2^{\alpha \beta \mu \nu}, \nonumber\\
T\gamma \gamma &:\,\, \frac{2i}{\Lambda} (f_1 c_w^2+f_2 s_w^2) K_1^{\alpha \beta \mu \nu}, \nonumber\\
T\gamma Z &:\,\, \frac{2i}{\Lambda} c_w s_w (f_2-f_1) K_1^{\alpha \beta \mu \nu}, \nonumber\\
Tgg &:\,\, \frac{2i f_9}{\Lambda} \delta^{a b} K_1^{\alpha \beta \mu \nu},
\label{Feynmanrules}
\end{align}
where $c_w$ and $s_w$ denote the cosine and sine of the Weinberg angle,
$v$ is the vacuum expectation value of the Higgs field and the two different tensor structures are given by
\begin{align}
K_1^{\alpha \beta \mu \nu}&=p_1^\nu \, p_2^\mu \, g^{\alpha \beta}- p_1^\beta \, p_2^\nu \, g^{\alpha \mu}-
p_2^\alpha \, p_1^\nu \, g^{\beta \mu}+p_1 \cdot p_2 \, g^{\alpha \nu} g^{\beta \mu},\nonumber\\
K_2^{\alpha \beta \mu \nu}&=g^{\alpha \nu} g^{\beta \mu}.
\label{tensorstructures}
\end{align}
The indices $\mu$ and $\nu$ correspond to the \spin2 field (which is
symmetric in the Lorentz indices), $\alpha$ is the index of the first gauge boson, whose
incoming four-momentum is denoted as $p_1$ and $\beta$ is the index of
the second one with four-momentum $p_2$. 
$a$ and $b$ are the color indices of the two gluons. 
The propagator of the \spin2 field is the same as in
Ref.~\cite{Frank:2012wh}, yet enlarged by an additional gluonic
contribution to the decay width.

% The propagator of the \spin2
% field with momentum $k$, i.e.\ the Fourier transform of 
% $\left<0\left|\mathcal{T}\left(T^{\mu\nu}(x)T^{\alpha\beta}(0)\right)\right|0\right>$, 
% is given by~\cite{Hagiwara:2008jb,Giudice:1998ck}
% \begin{equation}
% \frac{i B^{\mu \nu\alpha\beta}(k)}{k^2-m_T^2+im_T\Gamma_T}, \label{propagator1}
% \end{equation}\\
% where $m_T$ is the mass of the \spin2 particle, $\Gamma_T$ is its width 
% and $B^{\mu \nu\alpha\beta}(k)$ is defined as
% \begin{align}
% B^{\mu \nu\alpha\beta}(k)&=\frac{1}{2}\left(g^{\mu\alpha}g^{\nu\beta}+g^{\mu\beta}g^{\nu\alpha}-g^{\mu\nu}g^{\alpha\beta}\right)
% +\frac{1}{6}\left(g^{\mu\nu}+\frac{2}{m_T^2}k^\mu k^\nu\right)\left(g^{\alpha\beta}+\frac{2}{m_T^2}k^\alpha k^\beta \right)\nonumber\\
% &\quad -\frac{1}{2m_T^2}\left(g^{\mu\alpha}k^\nu k^\beta+g^{\nu\beta}k^\mu k^\alpha+g^{\mu\beta}k^\nu k^\alpha+g^{\nu\alpha}k^\mu k^\beta\right).
% \label{propagator2}
% \end{align}
% Explicit expressions for the partial decay widths into all possible
% final states are given in App.~\ref{sec:decaywidths}.

Since the present \spin2 model is based on an effective Lagrangian
approach, it violates unitarity above a certain energy scale. In order
to parametrize high-energy contributions beyond this effective model, we 
use a formfactor, which is multiplied with the
amplitudes:
\begin{equation}
F_{\text{Spin-2}}=\left( \frac{\Lambda_{\ff}^2}{|p_1^2|+\Lambda_{\ff}^2}
\cdot \frac{\Lambda_{\ff}^2}{|p_2^2|+\Lambda_{\ff}^2} \cdot 
\frac{\Lambda_{\ff}^2}{|k_{\text{sp2}}^2|+\Lambda_{\ff}^2} \right)^{n_{\ff}}. \label{formfactor}
\end{equation}
Here, $p_1^2$ and $p_2^2$ are the squared invariant masses of the initial
gauge bosons and $k_{\text{sp2}}^2$ is the squared invariant mass of an $s$-channel
\spin2 particle. The energy scale $\Lambda_{\ff}$ and the exponent
$n_{\ff}$ are free parameters, describing the scale of the
cutoff and the suppression power, respectively.\\
\\
Anomalous couplings of a Higgs boson to electroweak bosons can also be described by an effective Lagrangian 
approach~\cite{anomHiggscoupl, hep-ph/0403297, Arnold:2011wj}:

\begin{align}
\label{spin0Lagrangian}
 \mathcal{L}_{\text{Spin-0}} = \frac{1}{\Lambda_5} H & \left(g_{5e}^{HWW} W^{+}_{\mu \nu} W_{-}^{\mu \nu} + g_{5o}^{HWW} \, \widetilde{W}^{+}_{\mu \nu} W_{-}^{\mu \nu} + \frac{g_{5e}^{HZZ}}{2} Z_{\mu \nu} Z^{\mu \nu} + \frac{g_{5o}^{HZZ}}{2} \widetilde{Z}_{\mu \nu} Z^{\mu \nu} \right. \nonumber \\
& \left. + \frac{g_{5e}^{H\gamma \gamma}}{2} A_{\mu \nu} A^{\mu \nu} + \frac{g_{5o}^{H\gamma \gamma}}{2} \widetilde{A}_{\mu \nu} A^{\mu \nu}  + g_{5e}^{HZ\gamma} Z_{\mu \nu} A^{\mu \nu} + g_{5o}^{HZ\gamma} \widetilde{Z}_{\mu \nu} A^{\mu \nu} \right).
\end{align}
$\widetilde{V}_{\mu \nu}$ are the dual field strength tensors $\widetilde{V}_{\mu \nu} = 
\frac{1}{2} \varepsilon_{\mu \nu \rho \sigma} V^{\rho \sigma}$, ${\Lambda_5}$ is the energy scale of the underlying new
physics and $g_{5e(o)}^{HVV}$ denote the free coupling parameters corresponding to $\mathcal{CP}$-even (-odd) operators. \\
Analogous to the \spin2 case, a formfactor can be multiplied with the vertices to modify the high-energy behavior:
\begin{equation}
F_{\text{Spin-0}} = \frac{\Lambda^2_{\ff\!_0}}{|p_1^2| +
\Lambda_{\ff\!_0}^2} \cdot \frac{\Lambda_{\ff\!_0}^2}{|p_2^2|
+\Lambda_{\ff\!_0}^2} \,,  \label{Spin0formfactor}
\end{equation}
with $\Lambda_{\ff\!_0}$ describing the energy scale of the cutoff.

\FloatBarrier\section{\label{sec:Calculation} {Elements of the Calculation}}

In the present analysis, we study the characteristics of Higgs and \spin2 resonances 
produced in gluon fusion and vector-boson fusion, which decay into 
$\gamma \gamma \,$, $W^+ W^- \rightarrow l^+ \nu l^- \bar{\nu}$ 
or $Z Z \rightarrow 4 l$. To this end, we use the parton-level Monte Carlo program
\textsc{Vbfnlo}\cite{Arnold:2011wj}.

\begin{figure}
% \vspace{1.5cm}
 \begin{minipage}{0.5\textwidth}%
		\includegraphics[width=0.95\textwidth]{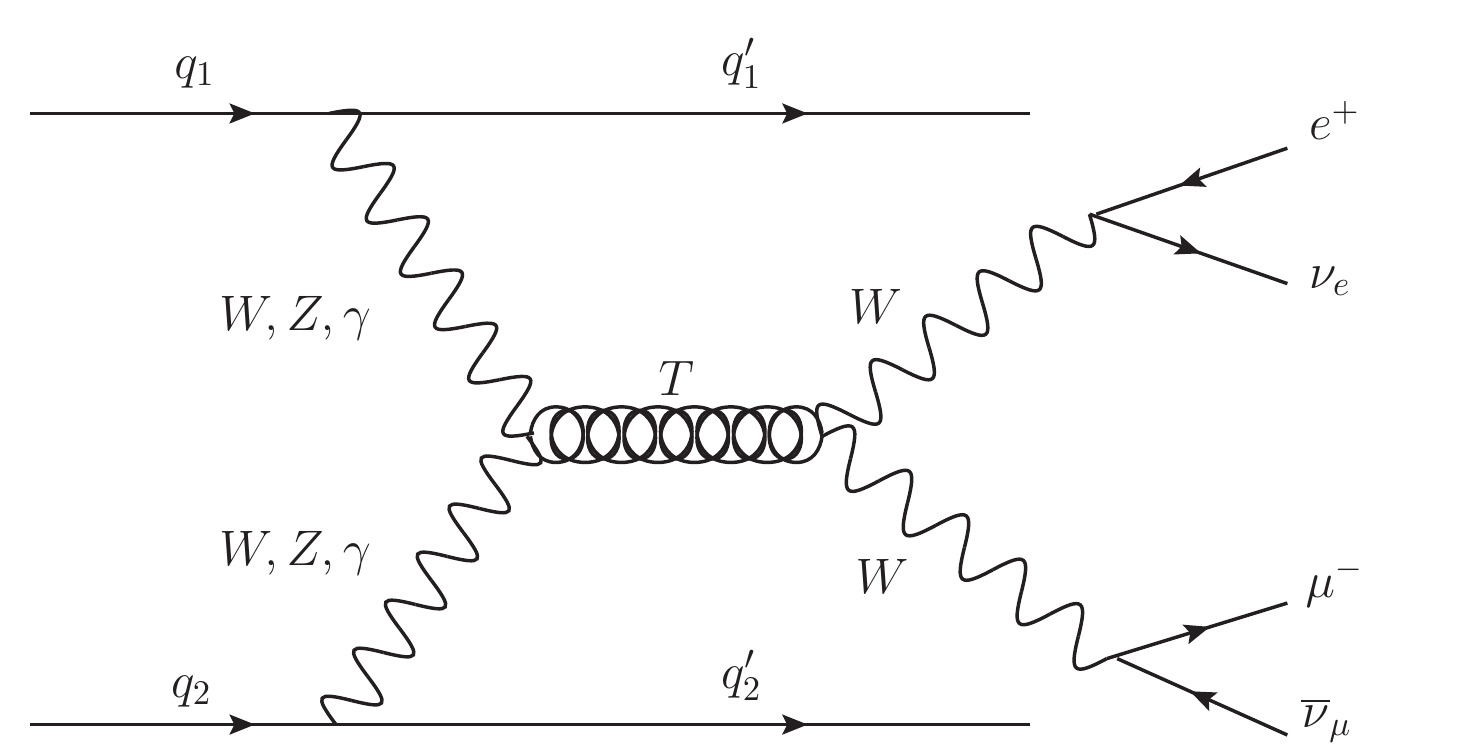}
	\end{minipage}
	\begin{minipage}{0.5\textwidth}%
		\includegraphics[width=0.95\textwidth]{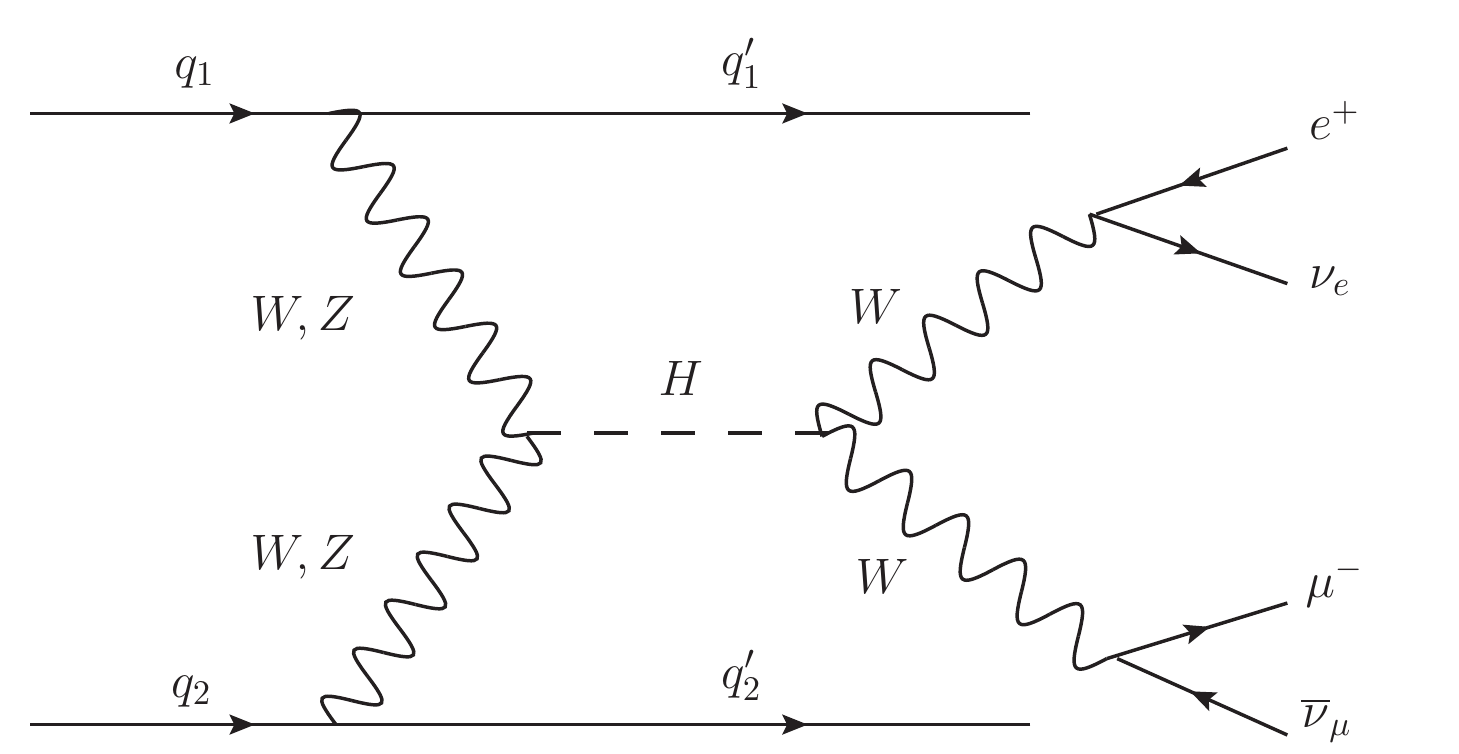}
	\end{minipage} 
% \vspace{-0.2cm}
\caption{Tree-level Feynman graphs of the VBF process $pp \rightarrow W^+ W^-\, jj \rightarrow e^+ \, \nu_e \,
\mu^- \, \overline{\nu}_\mu \, jj$. Left hand side: via a \spin2 resonance, 
right hand side: via a Higgs resonance.}
\label{figvbf}
\end{figure}

The analysis of Higgs and \spin2 resonances in vector-boson fusion is performed 
with NLO QCD accuracy. For the photon pair-production channel, our calculation 
is described in detail in Ref.~\cite{Frank:2012wh}, and we follow the same 
procedure also for the VBF $WW$ and $ZZ$ decay mode. In all cases, 
we only consider resonant diagrams, which are illustrated in Fig.\ref{figvbf} 
for the $WW$ channel at tree-level. Thereby, Higgs and \spin2 production are implemented 
as two separate processes in order to compare the characteristics of both cases. 
The SM continuum contributions are omitted, as interference effects are small 
due to the narrowness of the Higgs or \spin2 resonance. Since the
resonance is part of the electroweak sub-process, 
the NLO QCD corrections for the \spin2 case can be adapted from the 
existing calculation for VBF Higgs production~\cite{Figy:2003nv}. The real-emission contributions are  
obtained by attaching an external gluon to the two quarks lines of Fig.~\ref{figvbf} in all possible ways, 
which also comprises quark-gluon initiated sub-processes. 
Due to the color-singlet structure of VBF processes, the virtual corrections only comprise Feynman diagrams with a virtual 
gluon attached to a single quark line, which gives rise to vertex and quark self-energy corrections.

\begin{figure}
\centerline{\includegraphics[width=0.95\textwidth]{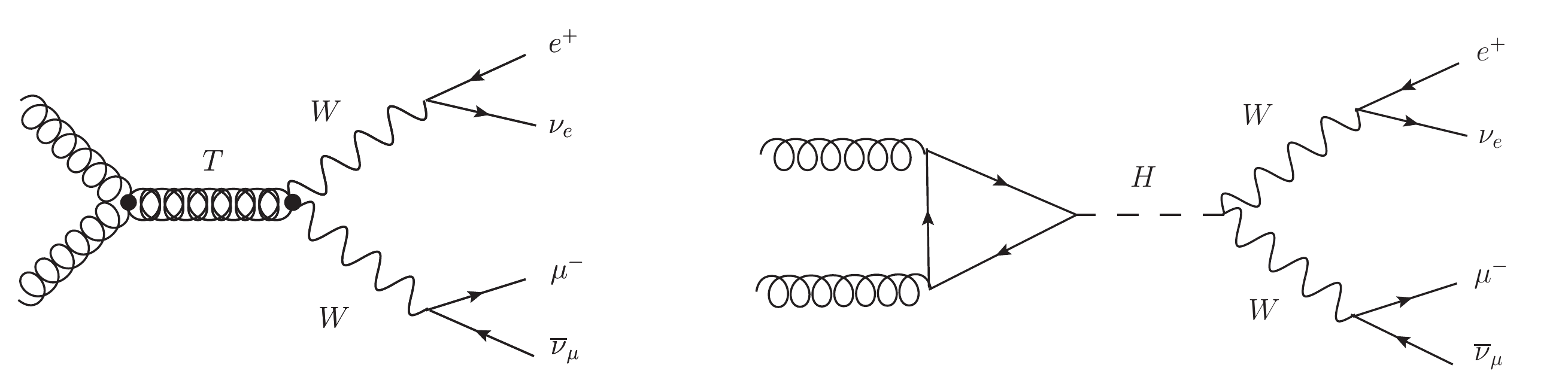}}
\caption{Feynman graphs of the process $gg \rightarrow W^+ W^- \rightarrow e^+ \, \nu_e \,
\mu^- \, \overline{\nu}_\mu$. \\ Left hand side: via a \spin2 resonance, 
right hand side: via a Higgs resonance.}
\label{figggww}
\end{figure}

Gluon-induced diboson-production processes are available in \textsc{Vbfnlo} 
at leading order, i.e. at the one-loop level for Higgs boson production, 
including anomalous Higgs couplings to electroweak gauge bosons for the decays. We have extended 
these implementations by \spin2-resonant processes in the effective Lagrangian approach, again omitting non-resonant 
diagrams. The contributing graphs are exemplified in Fig.~\ref{figggww} for $WW$ 
production. In the \spin2-resonant process $gg \rightarrow Z Z \rightarrow 4 l$, we also 
include intermediate virtual photons leading to a leptonic final state. 
To account for higher-order QCD corrections up to NNLL, which have sizable effects for Higgs 
production via gluon fusion~\cite{QCDcorr}, we multiply the LO cross sections calculated with 
\textsc{Vbfnlo} with a $K$-factor of 2.6 \footnote{This $K$-factor is rather high due to the scale choice 
$\mu_F= \mu_R = m_h$ for gluon fusion (see Sec.~\ref{sec:settings}). With $\mu_F= \mu_R = m_h/2$, it would 
be only $\approx 2.1$, because of a higher LO cross section.}, 
which was obtained by comparing with the value given in Ref.~\cite{LHCXSWG} 
(removing NLO EW corrections of about 5 \%~\cite{NLOEW} included therein). 
Thereby, we assume that higher-order QCD corrections for \spin2-resonant production in gluon fusion  
are the same as for Higgs production, since the operator structure of the $Tgg$ coupling, 
$T_{\mu \nu} G^{\alpha \nu}_{a} G^{a, \mu}_{\hspace{0.3 cm} \alpha}$, 
is analogous to the one of the effective $Hgg$ coupling, $H G^{\mu \nu}_{a} G^{a}_{\mu \nu}$. 
As higher-order QCD corrections also affect the decay of the \spin2 particle to gluons, 
we multiply the corresponding partial decay width with the $K$-factor 1.7, again following results obtained 
for the $H \rightarrow gg$ decay~\cite{decaycorr}. We note that only the assumed ratio of $K$-factors is 
relevant for \spin2 phenomenology, since the overall coupling strength of the \spin2 resonance to gluons, 
$f_9/\Lambda$, is a free parameter in our model.

% For the present analysis, we use the parton-level Monte Carlo program
% \textsc{Vbfnlo} \cite{Arnold:2011wj}, which simulates 
% vector-boson-fusion processes at hadron colliders with NLO QCD accuracy, 
% but also includes gluon-induced diboson-production processes at the leading order, 
% i.e. one-loop, level.

\subsection{\label{sec:settings}Input parameters and selection cuts}

As electroweak input parameters, we choose $m_W = 80.399$ GeV, $m_Z =
91.1876$ GeV and $G_F = 1.16637 \cdot 10^{-5} \,\text{GeV}^{-2}$, which
are taken from results of the Particle Data Group~\cite{PDG}. 
$\alpha$ and $\sin^2 \theta_W$ are derived from these quantities using
tree-level electroweak relations. 
We use the CTEQ6L1~\cite{Pumplin:2002vw} parton distribution functions
at LO and the CT10~\cite{Lai:2010vv} set at NLO with
$\alpha_s(m_Z)=0.118$.
In vector-boson-fusion processes, we set the factorization scale and the renormalization 
scale to $\mu_F= \mu_R = Q = \sqrt{|q_{if}^2|}$, where $q_{if}$ is the 4-momentum
transfer between the respective initial and final state quarks. 
% With this scale choice, LO calculations were found to give a good
% approximation of NLO cross sections and distributions, while
% the NLO results are hardly sensitive to the scale choice
% \cite{Bozzi:2007ur}. 
For gluon fusion or quark--antiquark-initiated diboson-production processes, 
we use a fixed scale of 126 GeV as 
factorization and renormalization scale.
Jets are recombined from the final state partons
by using the $k_{\bot}$ jet finding algorithm~\cite{Seymour:1997kj}. 
If not indicated otherwise, we consider a SM Higgs without anomalous $HVV$ couplings 
and a \spin2 resonance with couplings $f_1=0.04, f_2=0.08, f_5 = 10, f_9 = 0.04$ and 
$\Lambda = 6.4\, \text{TeV}$.
The parameters of the formfactor are $\Lambda_{\ff} = 400 \, \text{GeV},\,
n_{\ff} = 3$. These couplings produce rates which closely resemble those for the 
SM Higgs boson (see Table~\ref{crosssections}). 
The mass of the Higgs boson and the \spin2 particle is set to 126 GeV 
and we assume $pp$ collisions at a centre of mass energy of 8 TeV. 

Vector-boson-fusion events are characterized by two tagging
jets in the forward regions, with decay products of the vector bosons
lying in the central-rapidity region between them. 
By applying the following inclusive VBF cuts, these features can be used to 
improve the signal-to-background ratio in the VBF channels. 
The two tagging jets 
are supposed to lie inside the rapidity range accessible to the detector
and to have large transverse momenta: 
\begin{equation}
 p_{T,j}^{\text{tag}} > 30 \, \text{GeV}, \hspace{1cm} |\eta^\text{tag}_j| < 4.5.
\end{equation}
They are reconstructed from massless partons of pseudorapidity $|\eta| <
5$ and have to be well separated: 
\begin{equation}
\Delta R_{jj} \equiv \sqrt{(\eta_{j1}-\eta_{j2})^2 + (\phi_{j1}-\phi_{j2})^2} > 0.7.
\end{equation}
Due to the characteristic VBF kinematics, we require a large rapidity separation 
and a large invariant mass of the tagging jets,
\begin{equation}
\Delta \eta^\text{tag}_{jj} > 4, \hspace{1cm} m^\text{tag}_{jj} > 500 \, \text{GeV},
\end{equation}
which have to be located in opposite detector hemispheres, 
\begin{equation}
\eta_{j1}^{\text{tag}} \times \eta_{j2}^{\text{tag}} < 0.
\end{equation}
The charged decay leptons (or decay photons, respectively) are supposed 
to be located at central rapidities, to be well-separated from the
jets and to fall into the rapidity gap between the two tagging jets: 
\begin{equation}
 |\eta_l| < 2.5, \hspace{1cm} \Delta R_{lj} > 0.4, \hspace{1cm} \eta_{j,\text{min}}^{\text{tag}} < \eta_l < \eta_{j,\text{max}}^{\text{tag}}.
\end{equation}
Here, $l$ denotes a charged lepton or a photon, depending on the considered process.
In the leptonic decay channels, we apply a cut 
on the invariant mass of two oppositely charged leptons, 
\begin{equation}
m_{ll} > 15 \, \text{GeV}
\end{equation}
and require the transverse momentum of the charged leptons to be 
\begin{equation}
p_{T,l} > 10 \, \text{GeV} \hspace{0.3cm}
\text{in the $WW$ and} \hspace{0.3cm} p_{T,l} > 7 \, \text{GeV}  \hspace{0.3cm} \text{in the $ZZ$ mode.} 
\end{equation}
In the diphoton channel, we require 
\begin{equation}
p_{T,\gamma} > 20 \, \text{GeV}.
\end{equation}
In order to have isolated photons, we apply a minimal photon-photon \mbox{$R$-separation} 
\begin{equation}
\Delta R_{\gamma \gamma} > 0.4 
\end{equation}
and impose photon isolation from hadronic activity as suggested in 
Ref.~\cite{Frixione:1998jh}
with separation parameter $\delta_0=0.7$, efficiency
$\epsilon=1$ and exponent $n=1$.\\
%
%
% By imposing this set of cuts, the LO differential cross sections are
% finite, since they lead to finite scattering angles for the two jets. In
% the NLO calculation, initial state singularities appear, resulting from
% collinear quark and gluon splittings ($q \rightarrow q g$ and $g
% \rightarrow q \bar{q}$). 
% They are factorized into the PDFs. 
Divergences from
$t$-channel exchange of photons with low virtuality in real-emission
contributions are eliminated by imposing an additional cut on the photon
virtuality,
\begin{equation}
Q^2_\gamma > 4 \, \text{GeV}^2. 
\end{equation}
Analogous to Ref.~\cite{Oleari:2003tc}, the precise treatment of this
divergence does not appreciably affect the cross section, in particular
when VBF cuts are applied.\\
\\
In case of gluon fusion, we apply the same cuts on the charged 
decay leptons as in VBF, with  
\begin{equation}
p_{T,l} > 10 \, \text{GeV}, \hspace{1cm} |\eta_l| < 2.5, \hspace{1cm} m_{ll} > 15 \, \text{GeV}
\end{equation}
for the $W^+ W^- \rightarrow l^+ \nu l^- \bar{\nu}$ decay channel (and also for the diboson-production background) and
\begin{equation}
p_{T,l} > 7 \, \text{GeV}, \hspace{1.2cm} |\eta_l| < 2.5, \hspace{1cm}  m_{ll} > 15 \, \text{GeV}
\end{equation}
for $Z Z \rightarrow 4 l$. In the diphoton decay channel, we again require
\begin{equation}
p_{T,\gamma} > 20 \, \text{GeV}, \hspace{0.9cm} |\eta_\gamma| < 2.5, \hspace{1cm} \Delta R_{\gamma \gamma} > 0.4.
\end{equation}
In order to eliminate unwanted off-shell contributions in phase space regions where some of our approximations fail, 
we apply an additional cut on the invariant mass of all four final-state leptons (or the two photons, respectively) 
of $\pm 10$ GeV around the 126 GeV resonance in all gluon-fusion processes.

\FloatBarrier
\section{Results \label{sec:Results}}

In this section, we compare rates of a SM Higgs and a spin-2 resonance 
produced in VBF or gluon fusion for $\gamma \gamma \,$, $W^+ W^- \rightarrow 2 l 2 \nu$ 
and $Z Z \rightarrow 4 l$ decays. Focusing on the $WW$ decay channel, we present 
differential distributions which can be useful for a spin determination 
and study the impact of spin-0 and spin-2 model parameters and of next-to-leading order (NLO) QCD 
corrections in the VBF mode. In case of gluon fusion, distributions are determined at LO QCD 
and include a normalization factor $1/\sigma_{\text{LO}}$, while VBF figures are normalized to the 
NLO cross section. 
Due to the free coupling parameters $f_i$ of the spin-2
Lagrangian~(\ref{spin2lagrangian}), 
cross sections can be tuned such that they mimic those of a SM Higgs
within experimental and theoretical uncertainties. This was already 
shown for VBF photon-pair production in Ref.~\cite{Frank:2012wh}, yet is not only 
possible for single production and decay modes, but simultaneously for all the 
channels studied here. In case of a SM Higgs, the decay to two photons is suppressed 
compared to $WW$ and $ZZ$ decays, since the $H \gamma \gamma$ coupling is loop-induced. 
A similar suppression can be achieved in our model by tuning the different couplings $f_i$: 
As can be seen from the Feynman Rules in Eq.~\ref{Feynmanrules}, the coupling $f_5$ appears 
only in the $TWW$ and $TZZ$ vertex, but not in $T \gamma \gamma$ and $T \gamma Z$. By 
choosing $f_5 \gg f_1$, $f_2$ the decay to $\gamma \gamma$ can thus be suppressed compared 
to $WW$ and $ZZ$. 
That this is in fact possible for our parameter choice given above is illustrated in 
Table~\ref{crosssections}, which shows the integrated cross
sections for a SM Higgs and a \spin2 resonance. The statistical errors from the 
Monte Carlo integration are less than one per mill.
The NLO QCD corrections in the VBF channels are quite small for a Higgs and 
a \spin2 resonance, with $K$-factors
$K=\sigma_{\text{NLO}} / \sigma_{\text{LO}}$ between $1.01$ and $1.03$. 
Note that for graviton-like spin-2 models, it is not possible to 
obtain Higgs-like ratios in such a way \cite{EllisGraviton}. 
However, the ratio of Higgs and \spin2 rates depends on cuts, e.g. in the $WW$ channel, it 
changes significantly if additional upper cuts on the invariant dilepton mass and the azimuthal angle 
difference of the charged leptons are applied. With the ATLAS Higgs search cuts~\cite{HiggssearchATLAS} 
$m_{ll} < 50 \, \text{GeV}$ and $|\Delta\Phi_{ll}| < 1.8$, the SM Higgs cross section in 
$gg \rightarrow W^+W^- \rightarrow e^+ \, \nu_e \,\mu^- \, \overline{\nu}_\mu \,$ reduces from 30.1 fb 
(see Table~\ref{crosssections}) to 18.2 fb, whereas in case of \spin2, it is only 11.0 fb instead of 29.6 fb. 
This feature originates from the spin-dependent lepton kinematics in this channel, as we will discuss later. 
The width of the \spin2 resonance is far below the experimental resolution. 
With our default couplings, it is only about 5 keV.

\begin{table}
\begin{center}
\begin{tabular}{|c|c|c|c|}
\hline
Final State & Production mode & Higgs cross sec. [fb] & Spin-2 cross sec. [fb]\\
\hline\hline
$\gamma \gamma$ & VBF &  0.745 & 0.864\\
\cline{2-4}
& Gluon Fusion & 37.1 & 35.7\\
\hline\hline
$W^+W^- \rightarrow$ & VBF & 0.662 & 0.613\\
\cline{2-4}
$e^+ \, \nu_e \,
\mu^- \, \overline{\nu}_\mu \,$ & Gluon Fusion & 30.1 & 29.6\\
\hline\hline
$Z Z \rightarrow$ & VBF & $1.06 \cdot 10^{-2}$ & $0.982 \cdot 10^{-2}$\\
\cline{2-4}
 $ e^+ \, e^- \, \mu^+ \mu^-$ & Gluon Fusion & 0.468 & 0.446\\
\hline
\end{tabular}
\caption{Integrated cross sections for a SM Higgs and a spin-2 resonance 
with couplings \mbox{$f_1=0.04, f_2=0.08, f_5 = 10, f_9 = 0.04$} in VBF  
and gluon fusion (see text for details). 
The cuts of Section \ref{sec:settings} are applied.} \label{crosssections}
\end{center}
\end{table}

% In the following, when figures compare different values of coupling parameters,
% couplings $f_i$ which are not given explicitly are set to zero and
% $\Lambda$ is adjusted such that the cross section is approximately the 
% same as the one of the SM Higgs resonance. 
% If not indicated otherwise, differential distributions are determined 
% in the laboratory frame.

\begin{figure}
% \vspace{1.5cm}
 \begin{minipage}{0.5\textwidth}%
		\includegraphics[trim=30mm 20mm 50mm 55mm, width=\textwidth]{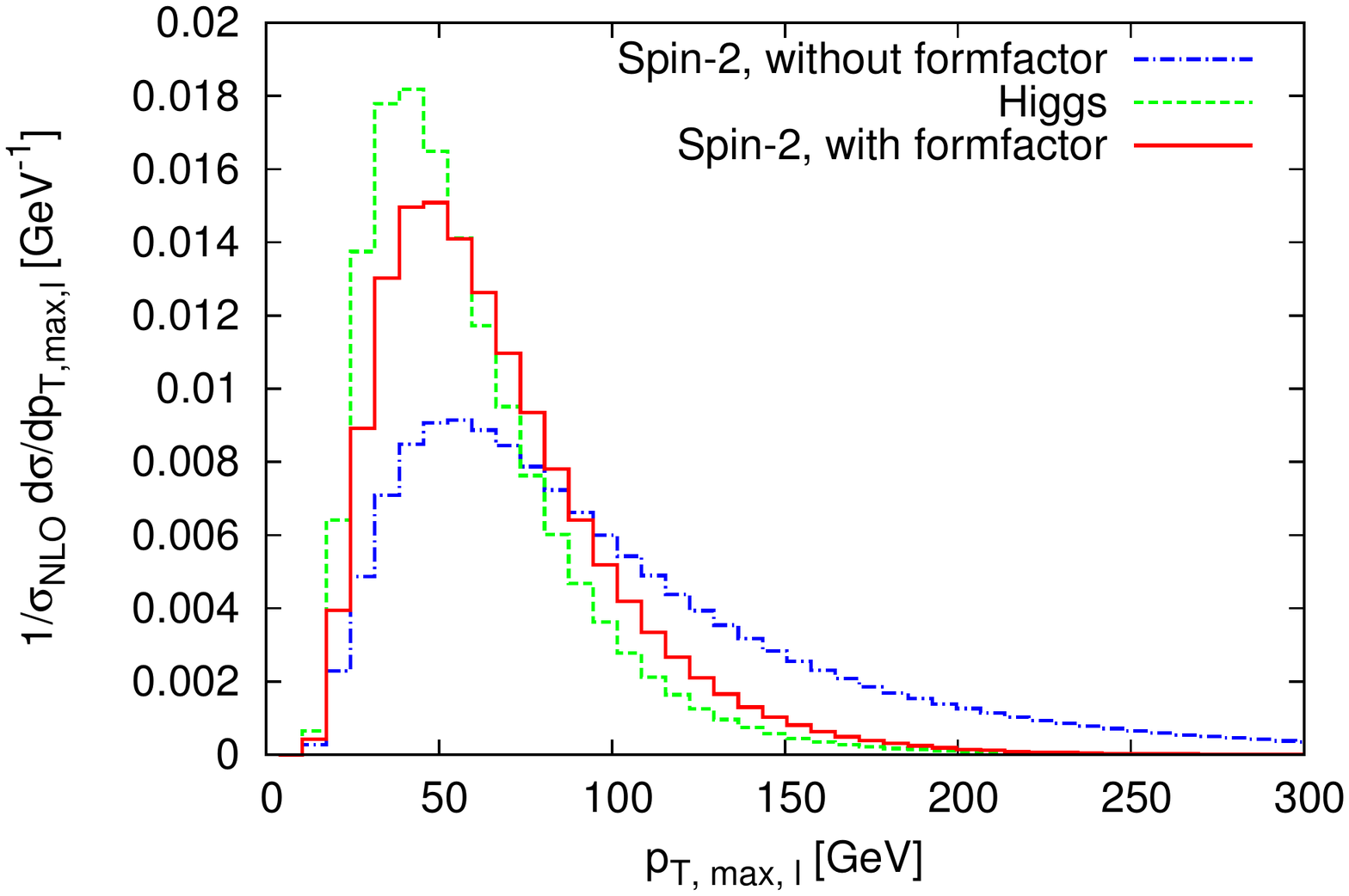}
	\end{minipage}
	\begin{minipage}{0.5\textwidth}%
		\includegraphics[trim=20mm 20mm 60mm 55mm, width=\textwidth]{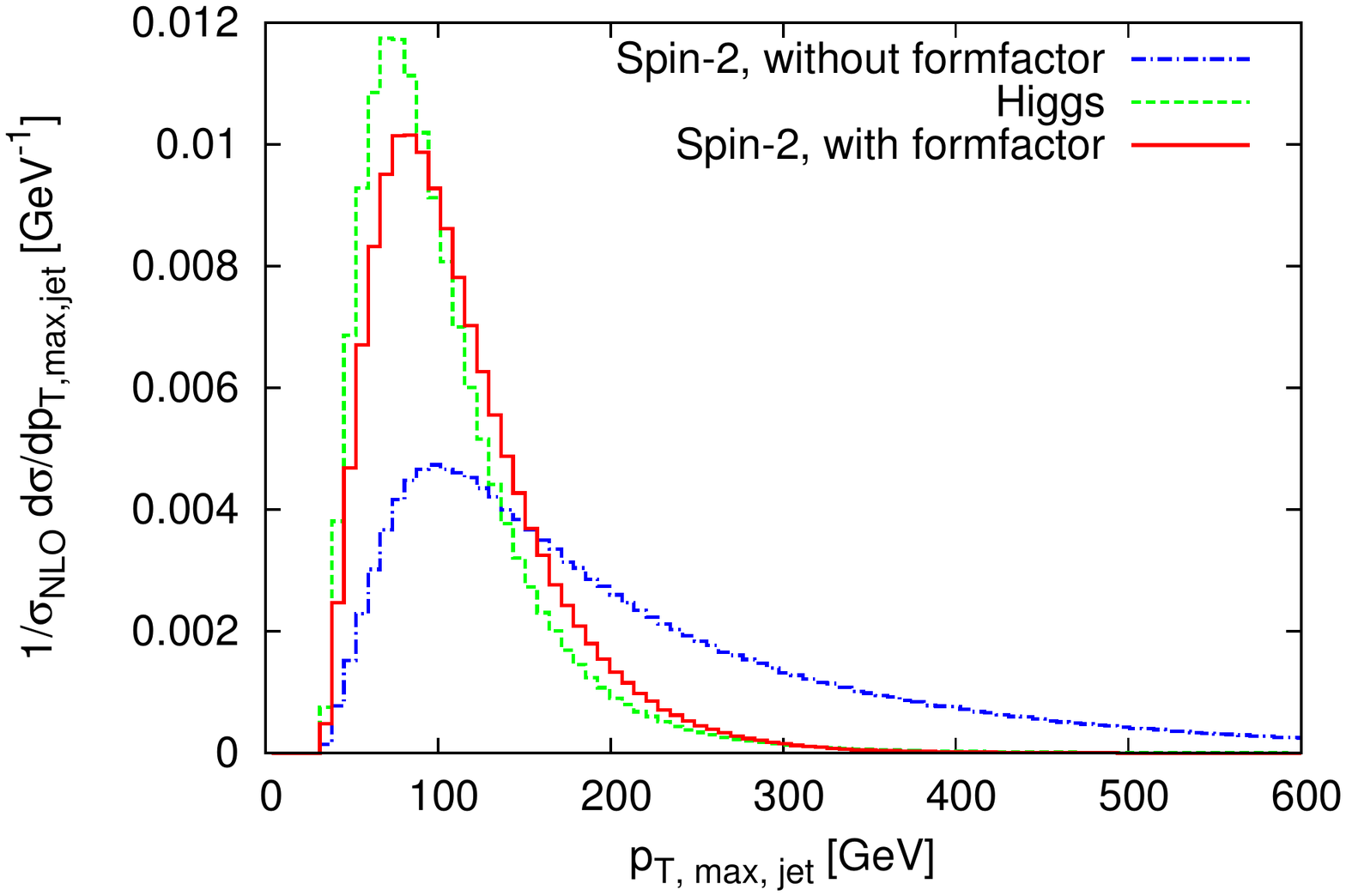}
	\end{minipage} 
% \vspace{-0.2cm}
\caption{Transverse-momentum distributions in VBF $W^+W^- \rightarrow e^+ \, \nu_e \,\mu^- \, \overline{\nu}_\mu \,$
 events for a SM Higgs and for a 
\spin2 resonance with couplings $f_1=0.04, f_2=0.08, f_5 = 10,$ \mbox{$f_9 = 0.04$,} 
with and without formfactor, 
 at NLO QCD accuracy. Left hand side: $p_T$ of the hardest final-state lepton, right hand side: 
$p_T$ of the tagging jet with the largest transverse momentum.} \label{pt}
\end{figure}

In Ref.~\cite{Frank:2012wh}, we have shown that in case of VBF
photon-pair production, not only cross sections, but also
transverse-momentum distributions of a \spin2 resonance can be adjusted
to those of the SM Higgs by choosing the \spin2 formfactor parameters of
Eq.~\ref{formfactor} to be $\Lambda_{\ff} = 400 \, \text{GeV},\,n_{\ff}
= 3$.
Again, this is simultaneously possible for $\gamma\gamma$, $WW$ and $ZZ$ decays 
within our set of formfactor parameters (see Fig.~\ref{pt} for $WW$). 
Therefore, transverse-momentum distributions are not sufficient for a spin determination; 
harder $p_T$ distributions for the \spin2 case without our specific formfactor setting 
originate from the higher energy dimensions of the couplings in the 
effective Lagrangian (\ref{spin2lagrangian}) instead of being an indicator of the spin. 
In fact, a similar behavior was found in Ref.~\cite{hep-ph/0403297} for a Higgs boson with anomalous  
couplings as in Eq.~\ref{spin0Lagrangian}.

\begin{figure}
% \vspace{1.5cm}
 \begin{minipage}{0.5\textwidth}%
		\includegraphics[trim=30mm 20mm 50mm 55mm, width=\textwidth]{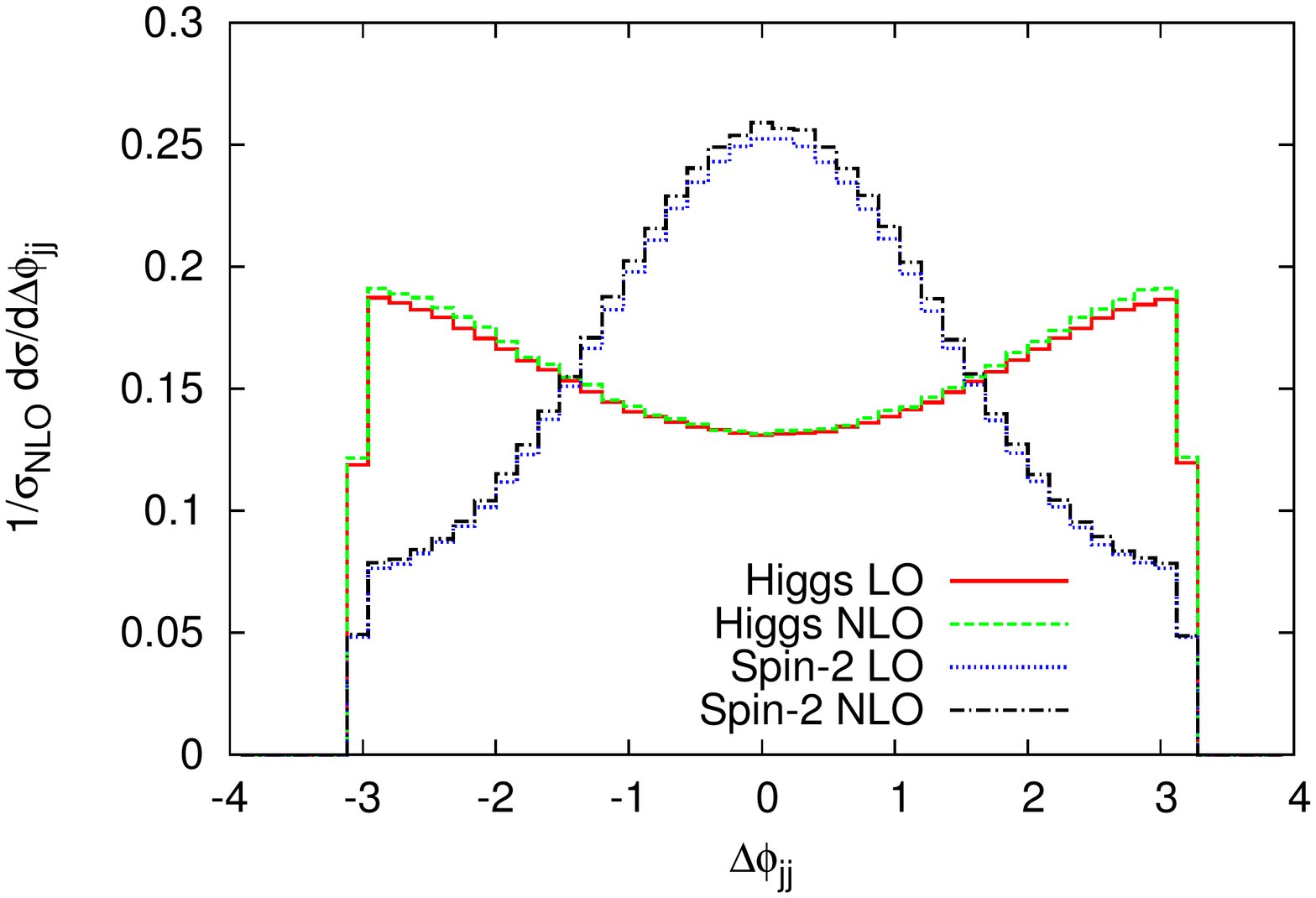}
	\end{minipage}
	\begin{minipage}{0.5\textwidth}%
		\includegraphics[trim=20mm 20mm 60mm 55mm, width=\textwidth]{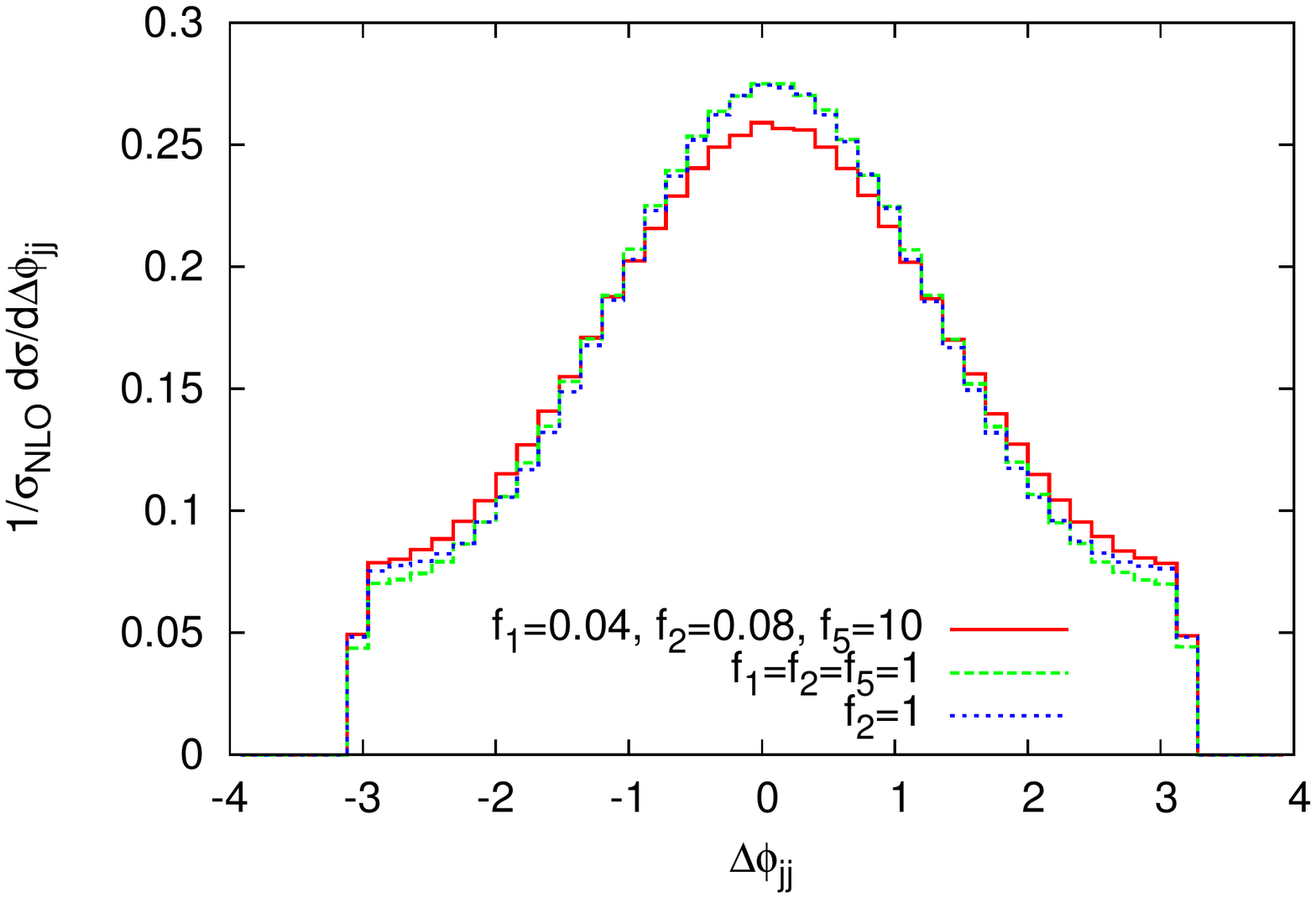}
	\end{minipage} 
% \vspace{-0.2cm}
\caption{Azimuthal angle difference of 
the two tagging jets for $W^+W^- \rightarrow e^+ \, \nu_e \,\mu^- \, \overline{\nu}_\mu \,$ in VBF. 
Left hand side: SM Higgs and \spin2 resonance with couplings $f_1=0.04, f_2=0.08, f_5 = 10, f_9 = 0.04$, 
both at LO and NLO QCD accuracy; right hand side: \spin2 resonance with different coupling parameters (always including $f_9 = 0.04$) at 
NLO QCD accuracy.} \label{deltaphijj}
\end{figure}

By contrast, the azimuthal angle difference between 
the two tagging jets was found to be an important variable for the determination 
of the spin  in VBF photon pair-production~\cite{Frank:2012wh}. 
This also holds for the $WW$ and $ZZ$ decay channels, with distributions similar to 
the diphoton case, as illustrated in Fig.~\ref{deltaphijj} for 
$W^+W^- \rightarrow e^+ \, \nu_e \,\mu^- \, \overline{\nu}_\mu \,$, including different \spin2 coupling parameters 
and the formfactor with $\Lambda_{\ff} = 400 \, \text{GeV},\,n_{\ff} = 3$. 
Note that the parameter choice $f_1=f_2=f_5=1$ resembles the electroweak part of the graviton scenario, 
but cannot reproduce observed Higgs rates, in contrast to our default choice.
Since the $\Delta\Phi_{jj}$ distribution features a clear difference 
between a SM Higgs and a \spin2 resonance, which
is nearly independent of the \spin2 couplings, the formfactor, the NLO QCD corrections and the decay mode, 
it is one of the most important tools to distinguish between \spin0 and \spin2 in VBF. 
However, the \spin0 distribution is model dependent: anomalous $HVV$
couplings (Eq.~\ref{spin0Lagrangian}) strongly alter the
$\Delta\Phi_{jj}$ distribution~\cite{hep-ph/0403297}. Furthermore, the
distribution of the SM Higgs depends on the cuts, e.g. with more stringent lepton $p_T$ cuts in the 
$WW$ or $ZZ$ mode, it is more central than the one of Fig.~\ref{deltaphijj}, which impairs the discriminating power.

\begin{figure}
% \vspace{1.5cm}
 \begin{minipage}{0.5\textwidth}%
		\includegraphics[trim=30mm 20mm 50mm 55mm, width=\textwidth]{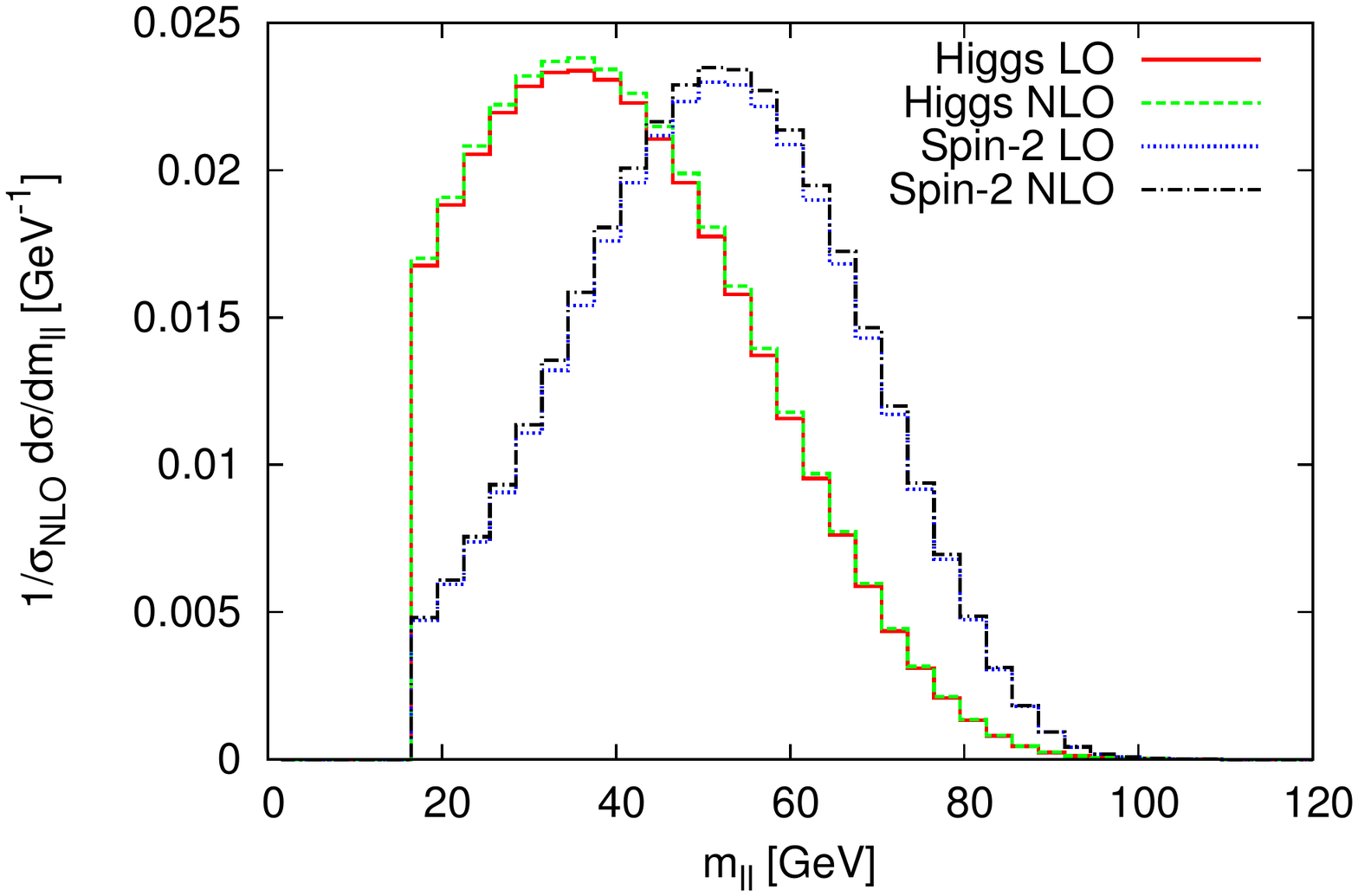}
	\end{minipage}
	\begin{minipage}{0.5\textwidth}%
		\includegraphics[trim=20mm 20mm 60mm 55mm, width=\textwidth]{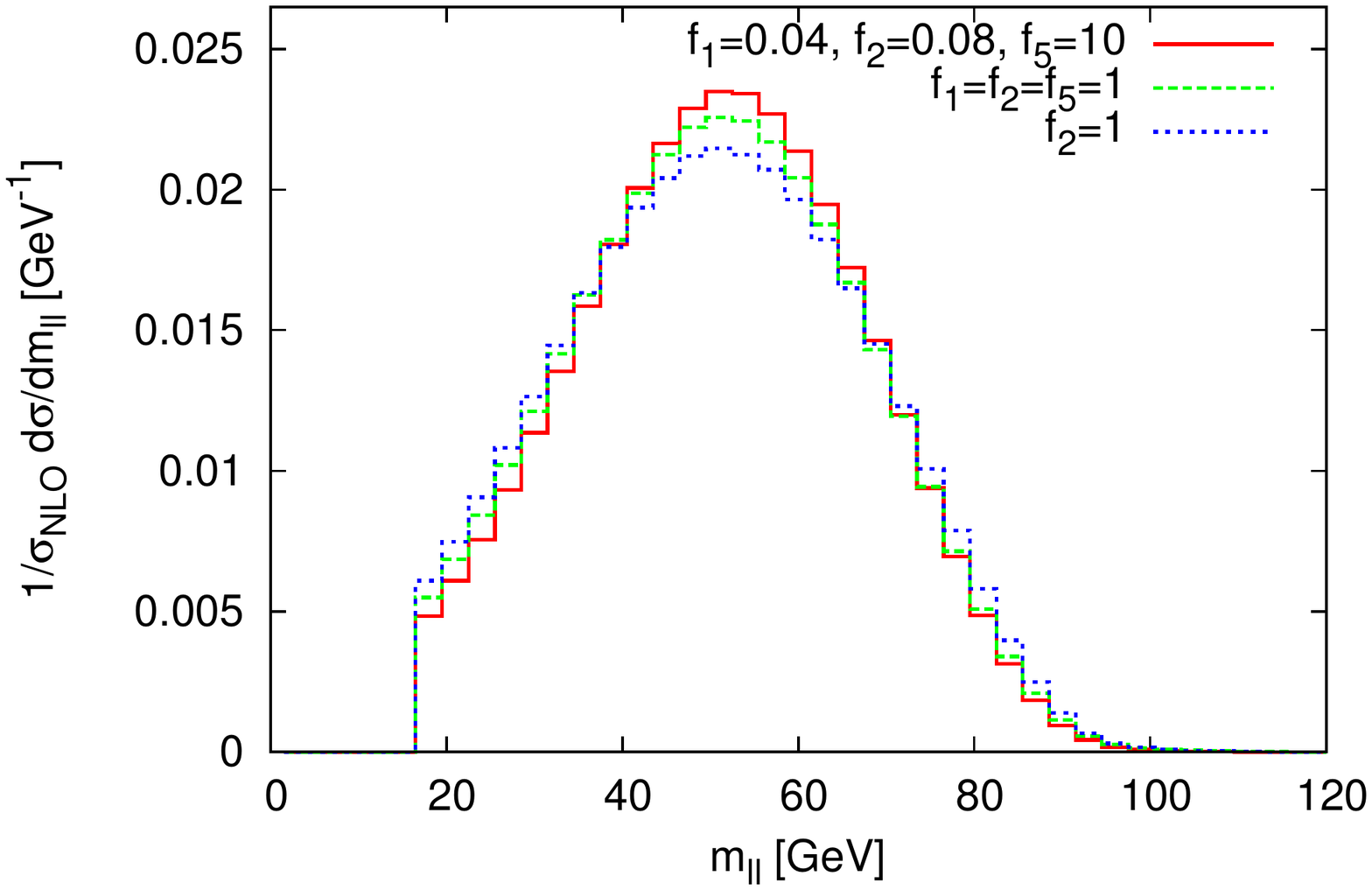}
	\end{minipage} 
% \vspace{-0.2cm}
\caption{Invariant mass of the two charged leptons for $W^+W^- \rightarrow e^+ \, \nu_e \,\mu^- \, \overline{\nu}_\mu \,$ in the VBF mode. 
Left hand side: SM Higgs and \spin2 resonance with couplings $f_1=0.04, f_2=0.08, f_5 = 10, f_9 = 0.04$, 
both at LO and NLO QCD accuracy; right hand side: \spin2 resonance with different coupling parameters (always including $f_9 = 0.04$) at 
NLO QCD accuracy. } \label{m2l_VBF}
\end{figure}

\begin{figure}
\vspace{1.5cm}
\centerline{\includegraphics[trim=25mm 10mm 55mm 70mm, width=0.5\textwidth]{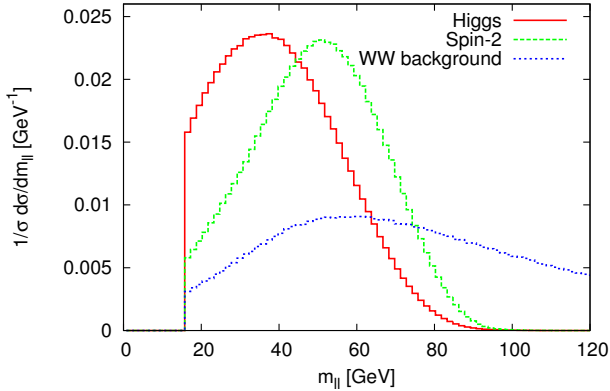}}
\vspace{-0.5cm}
\caption{Normalized distribution of the invariant dilepton mass for 
$gg \rightarrow W^+W^- \rightarrow e^+ \, \nu_e \,\mu^- \, \overline{\nu}_\mu \,$ 
 for a SM Higgs and a \spin2 resonance with couplings 
$f_1=0.04, f_2=0.08, f_5 = 10, f_9 = 0.04$ at LO QCD accuracy and the
diboson-production background including $q\bar{q} \rightarrow WW$ at NLO
QCD plus the continuum production diagrams of $gg \rightarrow WW$.}\label{m2l_gg_backgr}
\end{figure}

\begin{figure}
% \vspace{1.5cm}
 \begin{minipage}{0.5\textwidth}%
		\includegraphics[trim=30mm 20mm 50mm 55mm, width=\textwidth]{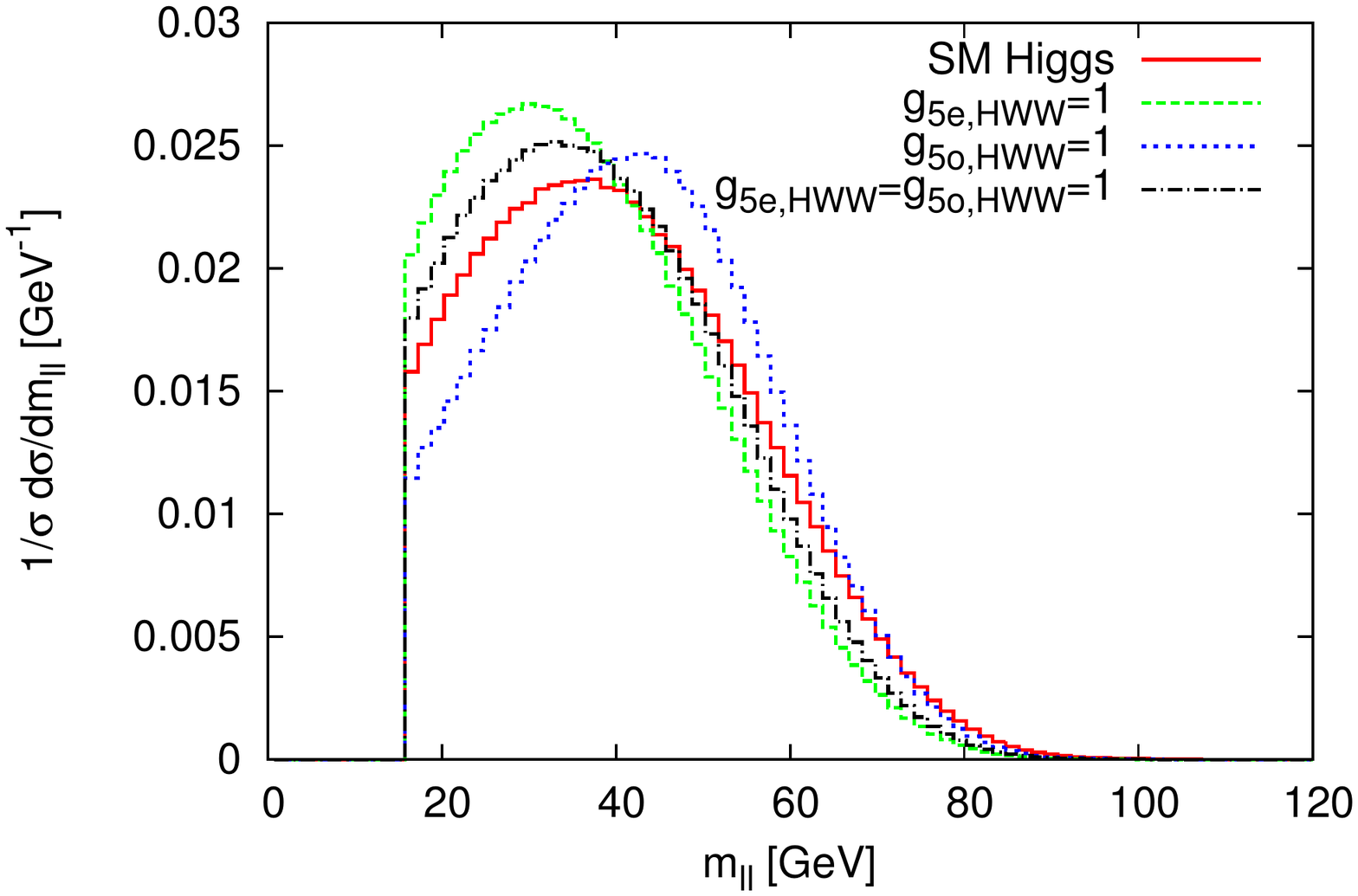}
	\end{minipage}
	\begin{minipage}{0.5\textwidth}%
		\includegraphics[trim=20mm 20mm 60mm 55mm, width=\textwidth]{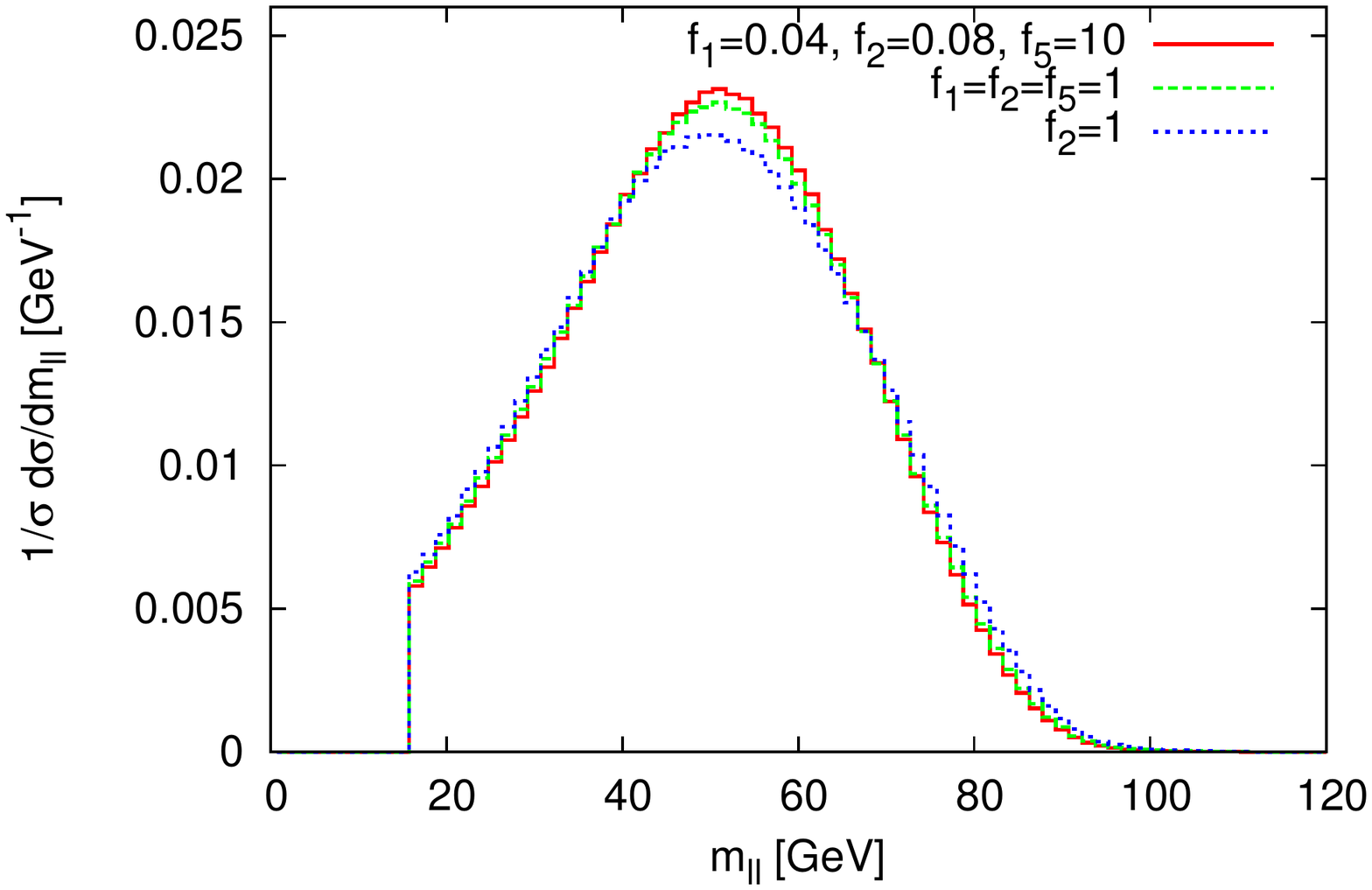}
	\end{minipage} 
% \vspace{-0.2cm}
\caption{Spin-0 and spin-2 model dependence of the invariant dilepton mass in the gluon-fusion mode (LO QCD accuracy). 
Left hand side: Higgs resonance with SM couplings, $\mathcal{CP}$-even and $\mathcal{CP}$-odd anomalous couplings; 
right hand side: \spin2 resonance with different coupling parameters (always including $f_9 = 0.04$).} \label{m2l_gg_coupl}
\end{figure}

In the $W^+ W^- \rightarrow l^+ \nu l^- \bar{\nu}$ decay channel, the invariant mass 
of the two charged leptons is another 
variable which is known to be an indicator of the spin~\cite{invdilepmass}. For a 
\spin0 resonance, the spins of the two $W$ bosons must be antiparallel, which leads to 
parallel momenta of the two charged leptons and therefore to a small invariant 
dilepton mass. Contrarily, in the \spin2 case, the spins of the $W$ bosons can be 
parallel, leading to antiparallel lepton momenta and a large invariant dilepton mass.\\
This is illustrated in Fig.~\ref{m2l_VBF} for the VBF mode, which shows that the invariant 
dilepton mass is much larger for a spin-2 resonance than for a SM Higgs and nearly independent 
of the spin-2 coupling parameters and the NLO QCD corrections. Note that these distributions 
include a cut $m_{ll} > 15 \, \text{GeV}$ 
(see Sec.~\ref{sec:Calculation}).\\
The same characteristic difference between a Higgs and a spin-2 resonance also arises in the 
gluon-fusion mode, which is depicted in Fig.~\ref{m2l_gg_backgr}. 
This figure additionally shows the normalized diboson-production background for comparison, including 
$q \bar{q} \rightarrow W^+W^- \rightarrow e^+ \, \nu_e \,\mu^- \, \overline{\nu}_\mu \,$ at NLO QCD accuracy and loop-induced 
$gg \rightarrow W^+W^- \rightarrow e^+ \, \nu_e \,\mu^- \, \overline{\nu}_\mu \,$ fermion-box contributions.
With an inclusive cross section of around 400 fb, this background exceeds the
one of a Higgs or spin-2 resonance significantly, even after placing more
stringent search cuts. Since the maximum of the invariant dilepton mass
distribution is nearly at the same position for the spin-2 signal and the
diboson continuum, a precise knowledge of the background is necessary.
In Fig.~\ref{m2l_gg_coupl}, the model dependence of the invariant dilepton mass distribution is studied for the spin-0 
and spin-2 case. As in the VBF mode (Fig.~\ref{m2l_VBF}), this observable is nearly independent of the spin-2 coupling parameters, whereas anomalous 
Higgs couplings can have a certain effect. Since only the $HWW$ couplings are relevant for the 
process $gg \rightarrow W^+W^-$, we only consider the first two terms of the Lagrangian~\ref{spin0Lagrangian} and we 
neglect the formfactor. Whereas the $\mathcal{CP}$-even coupling $g_{5e}^{HWW}$ alone (or the mixed case $g_{5e}^{HWW}=g_{5o}^{HWW}$) 
tend to shift the distribution to smaller values of $m_{ll}$, which facilitates the spin determination, the $m_{ll}$-distribution of a
$\mathcal{CP}$-odd Higgs with $g_{5o}^{HWW}$ is more similar to the one of a spin-2 resonance.

% Note that the invariant dilepton mass is also analyzed by ATLAS~\cite{ATLASWW} 
% and CMS~\cite{CMSWW}, yet not particularly for a spin determination.

% \begin{figure}
% % \vspace{1.5cm}
%  \begin{minipage}{0.5\textwidth}%
% 		\includegraphics[trim=30mm 20mm 70mm 50mm, width=\textwidth]{deltaphijj_f5_verschpara.pdf}
% 	\end{minipage}
% 	\begin{minipage}{0.5\textwidth}%
% 		\includegraphics[trim=20mm 20mm 80mm 50mm, width=\textwidth]{deltaphijj_trip_LHC8.pdf}
% 	\end{minipage} 
% % \vspace{-0.2cm}
% \caption{Azimuthal angle difference between the two tagging jets for a
% \spin2 resonance with different coupling parameters at NLO QCD accuracy. Left hand side: 
% \spin2 singlet, right hand side: \spin2 singlet and triplet.} \label{deltaphijj_verschpara}
% \end{figure}

%%%%%%%%%%%%%%%%%%%%%%%%%%%%%%%%%%%%%%%%%%%%%%%%%%%%%%%
\FloatBarrier\section{Conclusions \label{sec:conclusions}}
%%%%%%%%%%%%%%%%%%%%%%%%%%%%%%%%%%%%%%%%%%%%%%%%%%%%%%%

We have studied the characteristics of different \spin0 and \spin2 
hypotheses in order to determine the spin of the new resonance 
discovered at the LHC. To this end, we have implemented an effective model, 
describing the interaction of a \spin2 particle with SM gauge bosons, into 
the Monte Carlo program \textsc{Vbfnlo}. Comparing rates of 
\spin0 and \spin2 resonances produced in gluon fusion or vector-boson 
fusion in the decay modes 
$\gamma \gamma \,$, $W^+ W^- \rightarrow 2 l 2 \nu$ 
and $Z Z \rightarrow 4 l$, we find that with a suitable choice 
of model parameters, a \spin2 resonance can approximately reproduce  
SM Higgs rates in the main detection channels. 
Likewise, transverse-momentum distributions of a \spin2 resonance can be adjusted 
to those of a SM Higgs by tuning formfactor parameters, leaving angular and 
invariant-mass distributions for a spin determination. 
In the VBF production mode, we
found the azimuthal angle difference between the two tagging jets to be a very 
important variable to distinguish between \spin0 and \spin2. 
Its characteristics are nearly independent of \spin2 model parameters, NLO QCD corrections 
and decay mode.
Furthermore, in the $W^+ W^- \rightarrow l^+ \nu l^- \bar{\nu}$ decay, the invariant mass 
of the two charged leptons clearly distinguishes between \spin0 and \spin2 in VBF as well as 
in gluon fusion. 
Anomalous \spin0 scenarios, however, can lead to distributions which significantly differ from 
those of the SM Higgs. Therefore, it is important to carefully disentangle spin and $\mathcal{CP}$ 
properties of the new resonance. Since our default \spin2 model is largely compatible with 
present rate measurements at the LHC, we suggest that similar parametrizations, in particular 
$f_5 \gg f_1$, $f_2$, are used for further 
spin studies at the LHC as candidate \spin2 models.

%%%%%%%%%%%%%%%%%%%%%%%%%%%%%%%%%%%%%%%%%%%%%%%%%%%%%%%
\FloatBarrier\section*{Acknowledgments}
%%%%%%%%%%%%%%%%%%%%%%%%%%%%%%%%%%%%%%%%%%%%%%%%%%%%%%%
\noindent
This research was supported in part by the Deutsche
Forschungsgemeinschaft via the Sonderforschungsbereich/Transregio
SFB/TR-9 ``Computational Particle Physics'' and the Initiative and
Networking Fund of the Helmholtz Association, contract HA-101(``Physics at
the Terascale''). J.F.\ acknowledges support by the
"Landesgraduiertenf\"orderung" of the State of Baden-W\"urttemberg.

%%%%%%%%%%%%%%%%%%%%%%%%%%%%%%%%%%%%%%%%%%%%%%%%%%%%%%

\end{document}